\documentstyle[prd,aps,citesort,rsfs,amsbsy]{revtex} 
\newcommand{\beq}{\begin{equation}}
\newcommand{\eeq}{\end{equation}}
\newcommand{\bea}{\begin{eqnarray}}
\newcommand{\eea}{\end{eqnarray}}

\def\dslaux#1#2{\setbox0=\hbox{$#1{#2}$}
 \rlap{\hbox to \wd0{\hss$#1/$\hss}}\box0}
\let\sla=\dsl
\begin{document}

\title{One Interesting New Sum Rule \\ Extending Bjorken's to order 
$\boldsymbol{1/m_Q}$ \\ 
}
\author{A. Le Yaouanc$^{a}$, D. Melikhov$^{b}$\footnote{
Alexander-von-Humboldt fellow}, V. Mor\'enas$^{c}$, L. Oliver$^{a}$, 
O. P\`ene$^{a}$, and J.-C. Raynal$^{a}$}
\address{
${}^a$ Laboratoire de Physique Th\'eorique, Universit\'e de Paris XI, 
B\^atiment 211, 91405 Orsay Cedex, France\footnote{
Laboratoire associ\'e au Centre National de la Recherche Scientifique - 
URA D00063}\\
$^b$ Institut f\"ur Theoretische Physik, Universit\"at Heidelberg,
Philosophenweg 16, D-69120, Heidelberg, Germany\\
$^c$ Laboratoire de Physique Corpusculaire\\
Universit\'e Blaise Pascal - CNRS/IN2P3
F-63177 Aubi\`ere Cedex, France}
\maketitle

\begin{abstract}

We explicitly check quark-hadron duality to order  $(m_b-m_c)\Lambda/m_b^2$ for
$b \to c l\nu$ decays in the limit $m_b-m_c \ll m_b$ including ground state and
orbitally excited hadrons. Duality occurs thanks to a new sum rule which
expresses the subleading HQET form factor $\xi_3$  or,  in other notations,
 $a_+^{(1)}$ in
terms of the infinite mass limit form factors  and some level splittings. We also
demonstrate the sum rule,  which is not restricted to the condition $m_b-m_c \ll
m_b$, applying OPE to the longitudinal axial component of the hadronic tensor
without neglecting the $1/m_b$ subleading contributions to the form factors.  We
argue that this method should produce a new class of sum rules, depending
on the current, beyond Bjorken,
Voloshin and the known tower of higher moments. Applying OPE to the vector
currents we find another derivation of the Voloshin sum rule. From independent
results on $\xi_3$ we derive a sum rule which involves only the
$\tau_{1/2}^{(n)}$ and $\tau_{3/2}^{(n)}$ form factors and the corresponding
level splittings. The latter  strongly supports a theoretical evidence that the
$B$ semileptonic decay into narrow orbitally-excited  resonances dominates  over
the decay into the broad ones, in apparent contradiction with
some recent experiments. We discuss this issue.

\end{abstract}

\begin{flushright}
LPT-ORSAY 00/27\\ 
HD-THEP-00-12\\
PCCF RI 0005\\
\end{flushright}
\section{Introduction}

It is well known  \cite{ac2m2} that
quark-hadron duality is valid to a good accuracy in
$b$-quark decay and particularly in semileptonic decay.
A systematic study of the corrections to duality 
\cite{cgg,bsuv93,mw,fls,shifman} using the powerful tools of Operator Product
Expansion (OPE)  \cite{wilson} and Heavy Quark Effective Theory (HQET), in
particular Luke's theorem \cite{luke}, has demonstrated that the first
corrections to duality only appear at second order, namely $O(\Lambda^2/m_Q^2)$
where $\Lambda$ is for the QCD scale and $m_Q$ is  one of the heavy quark masses
($m_b$ or $m_c$). For simplicity we leave aside in this letter  the
$O(\alpha_s)$ radiative  corrections notwithstanding their manifest practical
relevance.

The OPE based proof is very elegant and circumvents the detailed calculation of
the relevant channels. Precisely  this feature has generated  some doubts or at least
some worries. First of all there is the experimental problem of the $\Lambda_b$
life time  which has not yet been understood within OPE framework. Second it has
been asked if OPE could not miss some subtle kinematical effects related with the
delay in the opening of different   decay channels \cite{isgur}. We have shown
\cite{nous}  in a non-relativistic model that the latter effect does not affect
the validity of duality. 
  
A numerical calculation of the sum over exclusive channels in the 't Hooft two
dimensional QCD model \cite{thooft} reported a presence of a duality-violating
$1/m_Q$ correction in the  total width \cite{gl}. Later the summation was
performed analytically in  the case of the massless light quark \cite{b-thooft}.
Agreement between  the OPE and the exact result was found in this case through
$1/m_Q^4$ order.

The ``miraculous'' conspiracy of exclusive decay channels   to add up to the
partonic result and its OPE corrections may be expressed in terms of sum rules
which the hadronic matrix elements must satisfy in QCD
\cite{bjorken,iw,voloshin,greg}.  OPE was  first explicitly used to
derive Bjorken sum rule in \cite{iw}. 

To leading order in $\Lambda/m_b$ Bjorken sum rule straightforwardly implies
quark hadron duality for the semileptonic widths (the differential and the total
widths). The suppression of  the $O(\Lambda/m_b)$ corrections is not so direct.
The authors of \cite{bgm} have done a thorough study of the exclusive
contributions of the ground state $D$ and $D^\ast$ mesons up to order
$O(\Lambda^2/m_b^2)$. They have chosen the Shifman Voloshin (SV) \cite{sv} limit,
$\Lambda \ll m_b-m_c \ll m_b$, which drastically simplifies the calculation, but
 did not consider the orbitally excited states,  and
therefore could not check the matching between the sum of exclusive channels and
the OPE prediction to the order  $O(\Lambda (m_b-m_c)/m_b^2)$.

  Our first motivation was precisely to complete this part and add
  the $L=1$ excited states in the sum of exclusive channels. 
  We will discuss in section \ref{sec:OPE} why we neglect other
  excitations.   

While performing this task we had a surprise. We found that a {\it new sum rule},
eq.~(\ref{new}),  was needed beyond Bjorken, Voloshin, and the known tower of
higher moment sum rules \cite{bjorken,iw,voloshin,greg} and we found  that this
new sum  rule could be demonstrated from OPE. 

We believe that other new sum rules can be derived
along the same line. When the form factors are
taken at leading order in $1/m_b$, OPE  applied to different components of 
the hadronic tensor, or to different operators,  always provides
 the unique series: Bjorken sum rule, Voloshin sum rule
and higher moments. But when the next to leading contribution to the form factors
is considered, no such unicity holds anymore. Changing the current operators
 in the OPE might lead to several other sum rules at order $1/m_b$.

In the following we will simplify our task as much as possible. We will neglect
 radiative corrections. We will also leave aside  terms of order 
$O(\Lambda^2/m_b^2)$, which implies that operators with higher dimension than
identity may be neglected in the OPE and consequently that the inclusive results
may be computed only via the partonic contribution.

In the next section we will show how the equality of partonic and inclusive
 widths to the desired order demands for a new sum rule. In section \ref{sec:OPE}
 we will derive the latter sum rule from OPE applied to the T-product
of currents.  Finally in section \ref{sec:tau} we show interesting 
phenomenological consequences of the sum rule. We then conclude.

\section{Inclusive semileptonic widths}
\label{largeurs}

We work in the SV limit \cite{sv}, i.e. we assume the following
hierarchy

\beq	
	\Lambda \ll \delta m \ll m_b \eeq
where $\delta m \equiv m_b-m_c$ and $\Lambda$ is any energy scale stemming
from QCD, for example  the hadron-quark mass difference  $\overline \Lambda
\equiv m_B-m_b  = m_D-m_c + O(1/m_b)$ or the excitation energy. 
	
From OPE \cite{bsuv93} one expects quark-hadron duality to be valid up to 
$O(\Lambda^2/m_b^2)$
corrections, i.e. in terms of the double expansion in
$\delta m/m_b$ and $\Lambda/m_b$, it should be valid to
all orders $(\delta m/m_b)^n$ and $(\delta m/m_b)^n\Lambda/m_b$.
In fact we will restrict ourselves to check duality up to order 
$(\delta m/m_b)^2$ and $\delta m \Lambda/m_b^2$. The terms 
of order $\delta m \Lambda/m_b^2$ will turn out to be the trickiest.
Of course, in the preceding sentences we mean orders as compared 
to the leading contribution. For example the inclusive semileptonic
width is of order $(\delta m)^5$, which implies that we will compute
it up to order $\overline \Lambda (\delta m)^6/m_b^2$. 
In this letter the symbol $\simeq$ will always refer to neglecting
higher orders than those just mentioned. 
From OPE the partonic semileptonic 
decay width should equate the explicit sum of the corresponding 
exclusive decay widths up to $O(\Lambda^2/m_b^2)$ terms, i.e.
\cite{bgm}:
\beq
\Gamma(\overline B \to X_c l \nu) = \Gamma(b\to c l \nu)
+ O(\Lambda^2/m_b^2)
\label{inclu}
\eeq
with the semileptonic partonic width 
\beq
\Gamma(b\to c l \nu)
 = 32 \,K \,(\delta m)^5\left[
\frac 2 5 - \frac 3 5 \frac{\delta m}{m_b} + \frac 9 {35}
\frac{(\delta m)^2}{m_b^2}\right] 
\label{partonic}
\eeq 
where
\beq
K = \frac {G_F^2}{192 \pi^3} |V_{cb}|^2
\eeq
Using $M_B\simeq m_b+\overline \Lambda$ and $\delta M
\equiv M_B-M_D \simeq \delta m$ we get 
\beq
\Gamma(\overline B \to X_c l \nu) \simeq 32\, K \,(\delta M)^5\left[
\frac 2 5 - \frac 3 5 \frac{\delta M}{M_B} + \frac 9 {35}
\frac{(\delta M)^2}{M_B^2} - \frac{21}{35} \frac {\overline \Lambda
\delta M}{M_B^2}\right] 
\label{inclu2}
\eeq 
The ground state contribution is 
\cite{bgm}
\beq
\Gamma(\overline B \to (D+D^\ast)\, l \nu) \simeq 32\, K\, (\delta M)^5\left[
\frac 2 5 - \frac 3 5 \frac{\delta M}{M_B} + \frac {11-8 \rho^2} {35}
\frac{(\delta M)^2}{M_B^2} - \frac{1}{10} \frac { a_+^{(1)}
\delta M}{M_B^2}\right] 
\label{ground}
\eeq 
Strictly speaking nothing compels $a_+^{(1)}$ to be real and we must 
read $\Re[a_+^{(1)}]$ everywhere in this letter instead of $a_+^{(1)}$ and 
$\Re[\xi_3]$ instead of $\xi_3$.
The contribution of the first orbitally excited states may be computed using
results in \cite{llsw}. We get
\beq
\Gamma(\overline B \to (D_1+D^\ast_2)\, l \nu) \simeq 32\, K\, 
|\tau_{3/2}(1)|^2\left[\frac {16}{35}\frac{(\delta M)^2}{M_B^2}
-\frac{56}{35}\frac { \Delta_{3/2}
\delta M}{M_B^2}\right] 
\label{3demis}
\eeq
for the states with total angular  momentum of the light 
quanta $j=3/2$ and 
 $\tau_{j}(w)$ are the infinite
mass limit form factors $B\to D^{\ast\ast}$ as defined in \cite{iw}.
In all this letter we use for any state $n$ the notation
\beq
\Delta_n = M_n- M_0, \label{Delta}
\eeq
where 0 refers to the ground state.
\beq
\Gamma(\overline B \to (D^{\ast}_{1}+D^\ast_0)\, l \nu) \simeq 32\, K\, 
|\tau_{1/2}(1)|^2\left[\frac {8}{35}\frac{(\delta M)^2}{M_B^2}
-\frac{49}{35}\frac {\Delta_{1/2}
\delta M}{M_B^2}\right] 
\label{1demi}
\eeq 
for the lowest $j=1/2$ states. 

To the order considered, quark-hadron duality of the semileptonic
decay widths implies the equality of the r.h.s. of eq. (\ref{inclu2}) 
with the sum of the r.h.s's of eqs (\ref{ground}), (\ref{3demis}) and 
(\ref{1demi}) to which
we need to add the $L=1$ radially excited states. Their contributions
are identical to eqs.~(\ref{3demis}) and (\ref{1demi}) with the replacement
$\tau_j \to \tau_j^{(n)}$ and $\Delta_j\to \Delta_j^{(n)}$. 
The terms proportional to $(\delta M/M_B)^2$ match thanks to Bjorken sum
rule \cite{bjorken,iw}:
\beq
\rho^2-\frac 1 4 = \sum_n \left[|\tau^{(n)}_{1/2}|^2 + 2 |\tau^{(n)}_{3/2}|^2
\right]\label{bjorken}
\eeq
From now on,
unless specified, it is understood that the form factors are taken at $w=1$.
Taking into account Voloshin sum rule \cite{voloshin}
\beq
\overline \Lambda = \sum_n \left[2\, \Delta^{(n)}_{1/2} \,|\tau^{(n)}_{1/2}|^2
+ 4\,\Delta^{(n)}_{3/2}\, |\tau^{(n)}_{3/2}|^2\right],
\label{voloshin}
\eeq
the matching of the terms of order $\Lambda \delta M/M_B^2$ leads to 
the requirement
\beq
a_+^{(1)}= 4\,\sum_n \left[\Delta^{(n)}_{1/2}\,|\tau^{(n)}_{1/2}|^2
- \Delta^{(n)}_{3/2}\, |\tau^{(n)}_{3/2}|^2\right]
\label{new}
\eeq

The sum rule (\ref{new}) is the main result of this paper. The 
preceding lines can be taken as a derivation of the sum rule, since we
simply have made explicit the result from OPE, eq. (\ref{inclu2}).
However, one might feel uncomfortable in view of the peculiarity
of the SV kinematics, one might fear that some exception to OPE could
happen there. Furthermore, as recalled in the introduction, OPE has
been repeatedly submitted to various interrogations. Therefore, we will
rederive in the next section the sum rule (\ref{new}) in a less 
questionable manner.

Let us note that in the {\it vector current case}, we do not need the $a_{+}$
form factor.  In that case, matching of  the $(\delta M/M_B)^2$ and
$\Lambda \delta M/M_B^2$ terms occurs thanks to Bjorken and Voloshin sum rule
only - or conversely we can invoke duality to demonstrate these sum rules. In
particular, it gives a demonstration of Voloshin sum rule just from the same
duality requirement invoked by Isgur and Wise to derive Bjorken sum rule:
 the Voloshin sum rule comes from the matching of $\Lambda \delta M/M_B^2$
terms. \par It is in the axial case or in the $V-A$ case (which corresponds to
the sum of vector and axial contribution) that we need the new sum rule. More
precisely, we can separate also the contributions with definite helicity of the
lepton pair. In the transverse helicity case, there is still matching from just
Bjorken and Voloshin sum rule. In fact the need for a new sum rule occurs in the
axial current and for {\it longitudinal} helicity. We obtain indeed for 
the $\lambda=0$ helicity of the axial current:
\beq
\Gamma(b \to c l \nu)_{A,\lambda=0}\simeq 4\, K \,(\delta M)^5\left[
\frac 4 3 - 2~ \frac{\delta M}{M_B} + \frac 4 {5}
\frac{(\delta M)^2}{M_B^2} - 2~ \frac {\overline \Lambda
\delta M}{M_B^2}\right] 
\eeq

\begin{eqnarray}
\Gamma(\overline B \to D^* l \nu)_{A,\lambda=0} \simeq 4\, K \,(\delta M)^5 &&\left[
\frac 4 3 - 2~ \frac{\delta M}{M_B} + (1-\frac 4 {5}~\rho^2)
\frac{(\delta M)^2}{M_B^2} -\frac{4}{5} \frac { a_+^{(1)}
\delta M}{M_B^2}\right] 
\\
\Gamma(\overline B \to D^{**} l \nu)_{A,\lambda=0} \simeq 4\, K \,(\delta M)^5
&&\left[
\frac 4 5 \sum_n \left[|\tau^{(n)}_{1/2}|^2 + 2 |\tau^{(n)}_{3/2}|^2
\right] \frac{(\delta M)^2}{M_B^2} \right. \nonumber\\
&&
-\frac {28} 5 \sum_n \left[ \Delta^{(n)}_{1/2} \,|\tau^{(n)}_{1/2}|^2
+ 2\,\Delta^{(n)}_{3/2}\, |\tau^{(n)}_{3/2}|^2\right] \frac {\delta M}{M_B^2}
\nonumber\\
&&\left.+\frac {24} 5 \sum_n \left[ \Delta^{(n)}_{1/2} \,|\tau^{(n)}_{1/2}|^2 \right]
\frac {\delta M} {M_B^2}\right]
\end{eqnarray}
whence we get the eq.~(\ref{new}) from the matching of ${\delta M}\over {M_B^2}$
terms.

\section{Derivation of the sum rule from OPE}
\label{sec:OPE}

The authors of \cite{blrw} have derived corrections to Bjorken and Voloshin sum
rules and to the resulting {\it inequalities} on $\rho^2$. We will follow the
same philosophy but including   the orbitally excited states in order to derive
$O(\Lambda/m_b)$ corrections, within our approximations,  to the {\it equalities}
resulting from the sum rules. We will use the differential semileptonic
distributions \cite{blok}.

Defining two currents which at present we take arbitrary:
\beq
J(x) \equiv \left(\overline b \Gamma c\right)(x),\quad J'(y) \equiv
\left(\overline c \Gamma' b\right)(y).
\label{currents}
\eeq
Their T product is
\beq
T(q) \equiv i \int d^4x e^{-i qx} <\overline B |T( J(x) J'(0))
|\overline B>
\label{Tdef}
\eeq
where the states are normalised according to
$<p | p'>= (2\pi)^3 \delta_3(\vec p'-\vec p)$.

Neglecting heavy quarks in the ``sea'', it is clear that $x<0$  receives
contributions from intermediate states with one $c$ quark and light quanta,
usually referred to as the direct channel, while $x>0$ receives contributions from
intermediate states with $b \overline c b$ quarks plus light quanta. This will be
referred to as the crossed channel, or $Z$ diagrams.  Expanding the r.h.s of
(\ref{Tdef}) on intermediate states $X$ in the $B$ rest frame, 
\beq
T= (2\pi)^3 \left [\sum_X \delta_3(\vec p_X+ \vec q)\frac 
{<\overline B |J(0)|X><X|J'(0)|\overline B>}{M_B-q_0-E_X}
-\sum_{X'} \delta_3(\vec p_{X'}- \vec q)\frac{<\overline B \overline X'|J(0)|0>
<0|J'(0)|\overline X'\overline B>}
{M_B+q_0-(E_{X'}+2M_B)}\right]
\label{Texp}
\eeq
where $X, X'$ are charmed states. Let us call ${\cal V}$ the typical
virtuality of the direct channels, $M_B-q_0-E_X\simeq {\cal V}$,  we will
take $q_0$ such that $\Lambda \ll {\cal V} \ll M_B$.  While the direct
channels ($X$) contribute like $1/{\cal V}$ to (\ref{Texp}), the crossed
channels ($X'$) contribute like $1/(m_D + {\cal V})$. In both cases the
denominator  is $\gg \Lambda$, which allows to use the leading contribution  to
OPE:
\beq
T=i \int d^4x e^{-i qx} <\overline B|\overline b(x) \Gamma S_c(x,0)
\Gamma' b(0) |\overline B> + \,O(1/m_c^2)
\label{OPE}
\eeq
where $S_c(x,0)$ is the free charmed quark propagator as long as $O(\alpha_s)$
corrections are neglected.  Assuming as usual that the $b$ quark has a momentum 
$p_b=m_b v + k$ with $k_\mu = O(\Lambda)$, the charmed quark propagator in
(\ref{OPE}) has two terms,  the positive energy pole  with a denominator $m_b v_0
+ k_0 -q_0 -E_c \simeq {\cal V}$ and the negative energy one with a 
denominator $m_b v_0 + k_0 -q_0 +E_c \simeq m_c + {\cal V}$.  Varying
${\cal V}$ independently of $m_b\simeq m_c$ one can check 
 that the direct channels sum up to the  contribution of the positive
energy pole of the charmed quark propagator.

As a result, considering now only resonances among the states $X$
and fixing $\vec q$ in the following, one gets equating the residues
\beq
\sum_n <\overline B |J(0)|n><n|J'(0)|\overline B>=
<\overline B |\overline b \Gamma \frac {\sla v'_q + 1}
{2 v_0'}\Gamma'b|\overline B>\label{residus}
\eeq  
where all the three-momenta are equal to $-\vec q$ in the $B$ rest frame 
and 
\beq
v'_q=\frac 1 {m_c} (-\vec q, \sqrt{\vec q^{\,2} + m_c^2})\label{vprime}
\eeq 

It is well known \cite{iw} that to leading order this leads to Bjorken
sum rule. Considering successive moments, i.e.
multiplying $T$ in (\ref{Tdef}) by $(q_0-E_0)^n$ ($E_0$ 
being the ground state energy) leads to a 
 tower of sum rules \cite{greg}, Voloshin sum rule when $n=1$, etc.

In the following {\it we will  stick to the $n=0$ moment}, but include the $1/m_b$
correction to the residues. Let us insist on this point. One may discover a tower
of sum rules by keeping the  form factors to leading order but considering
successive moments \cite{greg}. One may also discover new sum rules by sticking
to the lowest moment but considering the higher orders in the form factors. This
is not equivalent and leads to different sum rules, the first moment  yields
Voloshin sum rule eq~(\ref{voloshin}), the second  adds at least one new sum
rule, (\ref{new}), as we shall demonstrate now.  The distinction is important
since in practice both sum rules apply  to the same order in $1/m_b$. A
significant difference between the two types  of subleading sum rules is the
following: All the currents provide via OPE the same Voloshin sum rule because  
the form factors are all related by the heavy quark symmetry. On the contrary,
 when
the form factors  are taken at subleading order in $1/m_b$, different currents
have different corrective terms depending on several independent form factors,
 and OPE should yield different subleading sum rules. 
 In this letter we only consider
eq.~(\ref{new}) for its physical relevance, leaving other sum rules for a
forthcoming study.  

We now apply eq.~(\ref{residus}) with $J, J'$ substituted by  the vector current
$V^\mu$ and the axial one $A^\mu$. One may check that eq.~(\ref{residus})
applied to currents  projected perpendicularly to the $v,v'$ plane is trivially
satisfied,  including the $O(\Lambda/m_b)$ order, by Bjorken sum rule. Let us now
consider the vector current projected on the  $B$ meson four velocity: $V\cdot
v$. Among the orbitally excited states only the $J=1$ states contribute to the
wanted order. Dividing both sides of  eq~(\ref{residus}) by $(1+w)/(2v_0 v'_0)$ 
one gets using the results  of \cite{llsw} and \cite{fn} 
\beq
\frac {1+w}2 |\xi(w)|^2 + \sum_n (w-1)\left \{ 2 |\tau_{1/2}^{(n)}|^2
\left[1+\frac {\Delta^{(n)}_{1/2}}{m_b}\right] + (w+1)^2 |\tau_{3/2}^{(n)}|^2
\left[1+\frac {\Delta_{3/2}^{(n)}}{m_b}\right]\right\}
\simeq 1+(w-1)\frac {\overline \Lambda}{m_b}
\label{vector}
\eeq 
where we have neglected higher powers of $(w-1)$ and of $\Lambda/m_b$ than 
the first\footnote{Remember that we take $\overline \Lambda \sim \Delta_j
\sim \Lambda$}. The l.h.s is found by a straightforward application of 
\cite{fn} for the ground state and of \cite{llsw} for the excited ones. 
The r.h.s yields $(1+w_q)/(1+w)$ which has been transformed according to: 
\beq
w_q\equiv v.v'_q \simeq w + \vec q^{\,2} \left [\frac 1 {2m_c^2}-\frac
1 {2M_D^2}\right] \simeq w+ \frac {(w^2-1) \overline \Lambda}{m_b}.
\label{wq}\eeq

The leading terms in eq.~(\ref{vector}) simply reproduce Bjorken sum rule as
expected \cite{iw}, while the $O(\Lambda/m_b)$ terms provide Voloshin sum rule. 
This is {\it another derivation of Voloshin sum rule} which does not use higher
momenta.

Analogously the axial current projected on the $D$ meson velocity $v'$, $A\cdot
v'$ gives, inserted in eq.~(\ref{residus}) and after dividing both sides by
$(w-1)/(2v_0v'_0)$,

\[
\frac {1+w}2 |\xi(w)|^2 - \frac 4 {m_b} \xi_3(w)\xi(w)
+ \sum_n \Bigg \{ \left[2(w-1)-\frac {6(w+1)\Delta^{(n)}_{1/2}}{m_b} 
\right]|\tau_{1/2}^{(n)}|^2 \]\beq
+ (w-1)(w+1)^2 |\tau_{3/2}^{(n)}|^2\Bigg\}
\simeq 1-(w+1)\frac {\overline \Lambda}{m_b}
\label{axialv}
\eeq
 where $\xi_3$ in the notations of \cite{fn} is equal to $- a_+^{(1)}/2$
 used in \cite{bgm}. The matching of the $1/m_b$ terms in eq. (\ref{axialv})
 leads to the sum rule
 \beq
 \overline \Lambda + a_+^{(1)}= L_4(1) =  + 6 \sum_n \Delta^{(n)}_{1/2}
 |\tau_{1/2}^{(n)}|^2
 \label{sr2}\eeq 
$L_4$ being defined according to \cite{fn}.
Eliminating $\overline \Lambda$ from eqs.~(\ref{sr2}) and (\ref{voloshin})
we are left with eq.~(\ref{new}). 

 We can check
this result by using the method for sum rules developed earlier by Bigi and the
Minnesota group \cite{bigi}, which relies on a systematic $1/m_Q$ expansion of
the moments of the Lorentz invariants of the imaginary part of the hadronic
tensor, $w_i$. From their equation (131), we read :

\beq \int dq^0 w_2^{AA}(q^0,\vec q^{\,2})\simeq \frac {m_b}{E_c} \label{lw2}
\eeq
the terms left over being the power corrections due to higher dimension
operators. Computing from \cite{bgm} and \cite{llsw} the hadronic contribution to
the same integral at $\vec q=0$ i.e. $w=1$, we get the equation (with
$r_0=M_D^*/M_B$, $r_{1/2,3/2}=M_{D_{1/2,3/2}^{**}}/M_B$):

 \bea \int dq^0 w_2^{AA}(q^0, \vec{q}^{\,2}=0)={1 \over r_0} \left\lbrace {f^2 \over 4
M_B^2 r_0^2} +{(1-r_0) f a_+ \over r_0 } \right\rbrace \nonumber \\ + 
\left\lbrace  
{(1-r_{3/2})^2 \over r_{3/2}^2}\left({k_{A_1}^2 \over 24}-{f_A^2 \over 4}
\right)+{1 \over 4
r_{1/2}^2} (\lbrack (1+r_{1/2})  g_+-(1-r_{1/2}) g_- \rbrack ^2-g_A^2) 
\right\rbrace  \label{rw2}
\eea
with all form factors taken at $w=1$, and with notations for the $L=1$ form
factors $g_+,g_-,g_A,f_A,k_{A_1}$ to be found in \cite{llsw}. 
A sum over the $L=1$ excitations is unedrstood. If we now work in
the SV limit, we see that we need $g_-,f_A,g_A,k_{A_1}$ only in the HQET limit,
i.e.  $\tau_{1/2,3/2}$, except for some algebraic factors; as for
$g_+$, it is subleading, but at $w=1$, it is expressible in terms of
$\tau_{1/2}$ and we do not need to know any of the new subleading form
factors. In the $L=1$ contributions, only the $g_+ g_-$ term remains. We finally
end with the equation :

\beq \frac {M_B}{M_D} - \frac {\delta M}{M_B^2} a_+^{(1)}  + 6
 \frac {\delta M }{M_B^2}\sum_n  \Delta^{(n)}_{1/2} | \tau_{1/2}^{(n)}|^2 
 \simeq \frac
{m_b}{m_c} \label{w2} \eeq
which leads directly to eq.~(\ref{sr2}). 

In the preceding calculations we have systematically neglected the contributions
from higher orbital excitations or $L=0$ radial excitations.  This can be
justified as follows.  The leading $B$ transition to radially excited $L=0$ final
states or to $L=2$ final states are suppressed by a factor $\vec q^{\,2}/m_b^2$
due to three facts: first,  the current operator is proportional at
leading order to the identity 
operator or to $\vec \sigma_b$\footnote{
The heavy quark spin may be factorised out thanks to HQS.}, second, the orthogonality of the wave functions implies vanishing
at $\vec q=0$ in the $B$ rest frame  and, third,  parity implies an even power in
$\vec q$. This suppression leads to the well known fact that these terms appear
in the Bjorken sum  rule or in the differential widths with a $(w-1)^2$ factor as
compared to the ground state contribution.  On the contrary the   
$O(\Lambda/m_b)$ contributions to the axial form factors  for the same type of
transitions are not suppressed as compared to the ground state because the
current operator is no more proportional to identity neither to $\vec \sigma_b$.
 For example the transition
to radially or orbitally excited  $J^P=1^-$ states other than the  $D^\ast$ are
in principle of the same order of magnitude than the  $\propto a_+^{(1)}$ terms
mentioned above. However, in this letter we have only considered the terms
$\propto a_+^{(1)}$ via crossed terms, i.e. via cross products of the leading
order terms with the $O(\Lambda/m_b)$ ones, because we have neglected all
$O(\Lambda^2/m_b^2)$ contributions. Hence we are left with a suppression  of a
factor $\vec q^{\,2}/m_b^2$ in the hadronic tensors or the differential widths,
i.e. a factor $(w-1)$ as compared to the corresponding ground state contribution
and we can consequently neglect the $L=0$ radial excitations and the $L=2$
orbital ones.  $L=3$ contributions are negligible simply because the total
angular momentum $J\ge 2$ again leads to $(w-1)$ factors resulting from angular
momentum conservation (D-waves).     All other operators which are already
negligible for the ground state and the $L=1$ states are even more so for higher
excitations.
  
Turning now to a comparison of our different demonstrations,
 we should note that it is not really unexpected
that we find consistent results according to three approaches: imposing duality
to the widths (section \ref{largeurs}), imposing duality to the tensors
as in eqs.~(\ref{vector}) and (\ref{axialv}) and finally to the invariant tensors
eqs.~(\ref{lw2}) and (\ref{rw2}). Indeed, at fixed $q^0$ and $\vec q$ there
 is a linear relation
between the tensor components and the invariant tensors. It is as well true that
the formula for the decay widths before integrating on the $q^0$ variable is, 
for fixed $q^0$ and $\vec q$, linear in the tensor components.

 We might worry about what happens when we apply duality to the sum of the
residues. Integration over $q^0$ leads to a sum of residues multiplied by
$\delta$ functions and the position of the poles is different for each term in
the sum and still different for the quark contribution. As a consequence the
projector  which projects out $w_2$ from the tensor residues is different for
each term since it depends on $q^0$.  Still this difference does not lead to a
collapse of the sum rule thanks to Voloshin sum rule and the tower of higher
momenta sum rules: one can expand the difference between the intervening projectors in
powers of $q^0$ and the resulting alteration to the sum rule vanishes. Exactly
the same happens  when one computes the decay widths with the real kinematics on
each term.

\section{Phenomenological consequences}
\label{sec:tau}

Eq.~(\ref{sr2}) is phenomenologically relevant as it
expresses the dominant correction to the
zero recoil differential $B\to D l\nu$ decay width as a function
 of leading form factors and level spacings. Indeed
\beq  
\frac {d\Gamma(B\to D l \nu)}{dw} \propto (w^2-1)^{3/2}
\left[1-2 \left(\frac 1{2m_b}+\frac 1{2m_c}\right)\frac{M_B-M_D}{M_B+M_D}
L_4(1)\right].
\eeq
 
On the other hand, we may combine our result with an independent estimate of
the form factor $\xi_3$ \cite{xi3} from QCD sum rules\footnote{The definitions of $\xi_3$ differ by a factor $\overline
\Lambda$ in \cite{fn} and \cite{xi3}. We use the notations of \cite{fn}.}:  
\beq
\frac{\xi_3(1)}{\overline \Lambda } = \frac 1 3 + O(\alpha_s) = 0.6 \pm 0.2,\quad 
\frac{a_+^{(1)}}{\overline \Lambda } = -\frac 2 3 - O(\alpha_s) = - 1.2 \pm 0.4.
\label{xi_3}
\eeq
The dispersion formulation of the constituent quark model \cite{dmitri}
finds that $\xi_3(1)$ is 1/3 the average kinetic energy of the light quark.
For a light constituent mass of $m_u=0.25$ GeV it gives
\beq
\xi_3(1)= 0.17 \;{\rm GeV}, \qquad  \overline \Lambda = 0.5\; {\rm GeV}
\label{dm} 
\eeq
in perfect agreement with eq.~(\ref{xi_3}) for $\alpha_s=0$.

Combining (\ref{voloshin}), (\ref{new}) and (\ref{xi_3}), assuming
  $\alpha_s=0$ since we have neglected radiative corrections all 
  along this letter, we get 

\beq
{\sum_n \Delta^{(n)}_{1/2}\,|\tau_{1/2}^{(n)}|^2 \over
\sum_n\Delta^{(n)}_{3/2}\,|\tau_{3/2}^{(n)}|^2}={1\over 4}, 
\quad {\rm for} \quad \alpha_s=0
\label{taus}
\eeq
and
\beq
\sum_n \Delta^{(n)}_{1/2}\,|\tau_{1/2}^{(n)}|^2 = \frac 1{18}\, \overline
\Lambda,\qquad
\sum_n \Delta^{(n)}_{3/2}\,|\tau_{3/2}^{(n)}|^2 = \frac 2{9}\, \overline
\Lambda
\eeq
Notice that if we had, somehow inconsistently,
 taken $ \xi_3(1)/\overline \Lambda  = 0.6$ the result would not be qualitatively
 different. 
 
Since in all spectroscopic models the mass differences between the $j=1/2$ and
$j=3/2$ states turn out to be not so large, we conclude that the 
{\it  $\sum_n |\tau^{(n)}_{1/2}|^2$  
 are significantly smaller than the  $\sum_n |\tau^{(n)}_{3/2}|^2$} . 

Interestingly enough, this hierarchy $|\tau^{(0)}_{1/2}|^2 <
|\tau^{(0)}_{3/2}|^2$ was a clear outcome of a class of covariant quark models
\cite{vincent}. In \cite{vincent} four different potentials had been used within
the Bakamjian-Thomas covariant quark model framework. The potentials labeled ISGW,
VD, CCCN, and GI potentials in \cite{vincent} give  respectively for the ratio
$|\tau^{(0)}_{1/2}|^2/|\tau^{(0)}_{3/2}|^2$   the values 0.33, 0.09, 0.01 and
0.17. As a result these models predict a dominance of the $B\to D_{j=3/2} l\nu$ 
semileptonic decay widths by one order of magnitude over the $B\to D_{j=1/2}
l\nu$.  We will comment this prediction later. The same models \cite{vincent}
give for the l.h.s of eq.~(\ref{taus}) 0.39, 0.166, 0.151 and 0.247 respectively
for the ISGW, VD, CCCN, and GI potentials, in reasonable agreement with 1/4. It
 might not be mere luck if  the GI model, which fits the spectrum in the most
elaborate way, yields an almost too good agreement with the expectation 
(\ref{taus})\footnote{ We should nevertheless remember that the potentials used
in \cite{vincent} contain a Coulombic part which implies that some part of the
$O(\alpha_s)$ corrections might be implicit in these models.}. From
eq.~(\ref{xi_3}) we expect the r.h.s. of eq.~(\ref{new}) divided by that of
eq.~(\ref{voloshin}) to be close to $-2/3$. We have tested this with the numerical
calculations of
\cite{vincent}.  In all cases we find that the sums in the r.h.s of
eqs.~(\ref{voloshin})-(\ref{new}) saturate  very fast to their symptotic values.
At $n=3$ they are at less than 3\% in all cases. For the ratios
$a_+^{(1)}/\Lambda$ computed from the r.h.s of  eqs.~(\ref{voloshin})-(\ref{new})
one finds -0.51, -0.77, -0.79, -0.67 respectively for the ISGW, VD, CCCN, and GI
models.  This agreement with (\ref{xi_3}) is quite striking, and again GI is
embarrassingly good.

In more general terms, the prediction  \cite{vincent} that the $B$ meson decays
 dominantly into the narrow  resonances $j=3/2$  was comforted by a study within
 a constituent quark-meson model \cite{deandrea} as well as by a
 semi-relativistic study \cite{ebert}. A QCD sum rule analysis \cite{col}
 predicted rather a rough equality between these  form  factors contrarily to
 another one \cite{yuan} which concluded to an overwhelming dominance of the
 $j=3/2$ semileptonic decay  over the $j=1/2$.

It is fair to say that the general trend of theoretical models is 
to predict $3/2$ dominance and a total semileptonic  branching
ratio  into the orbitally excited states exceeding hardly 1 \%. 
It is well known that the $j=3/2$ are expected to be relatively 
narrow and are identified with the observed narrow resonances
$D_1(2422)$ and $D_2^\ast(2459)$. As far as the decay widths
 into the latter narrow resonances is considered, 
experimental results \cite{aleph} are  in rough agreement with
 \cite{vincent}    for the $B\to D_1(2422) l \nu$ and rather below
 \cite{vincent} for  $B\to D_2^\ast(2422) l \nu$. In brief, experiment
 is rather below the theoretical models for $B\to D_{3/2} l \nu$.
 The $j=1/2$ states are not easy to isolate, being very broad. 
 But thorough studies have been done of the channels $B\to D^{(\ast)} \pi
 l \nu$ and the resulting branching fraction is very large: $3.4 \pm 0.52 
 \pm 0.32 \%$ by DELPHI \cite{delphi} and $2.26 \pm 0.29 \pm 0.33 \%$ by
 ALEPH. 
 
 These experimental results are both welcome and puzzling. Welcome
 because these $B\to D^{(\ast)} \pi
 l \nu$ fill the gap between the inclusive semileptonic decay branching fraction
 of 10 - 11 \% and the sum $B\to (D + D^\ast) \,l \nu \simeq 7 \%$. They are 
 puzzling when one tries to understand which channels contribute to them. 
 As we have just said, the $j=3/2$ channels provide no more than 1 \%. 
 The remaining
 2 \% can come from the $j=1/2$,  from higher excitations or from a non-resonant
 continuum. Higher excitations are unlikely to contribute very much, being
 suppressed both by dynamics and phase space. In \cite{delphi} the 
 quoted $b\to D^{\ast\ast} l \nu$ branching fractions are very large, exceeding
 by far what is expected for example in \cite{vincent}. 
 
 The results presented in this letter are doubly relevant in the above
 discussion. First eq.~(\ref{taus}) seems to confirm the models which  find a
 dominance of the $3/2$ channels. Of course it is mathematically possible that
 that eq.~(\ref{taus}) is satisfied  while $|\tau_{1/2}^{(0)}| >
 |\tau_{3/2}^{(0)}|$, the higher excitations compensating for the sum rule. 
 Admittedly such a situation would look rather queer, and  as mentioned above ,
 the models \cite{vincent}, which agree rather well  with the new sum rule 
 eq.~(\ref{new}), also yield $|\tau_{1/2}^{(0)}|^2 < 0.35 |\tau_{3/2}^{(0)}|^2$.

 It is then hard to understand how
 the $b\to D^{\ast\ast} l \nu$ branching fractions can be as large as quoted
 in \cite{delphi} in view of the smallness of the experimental  $B\to D_{3/2}
 l\nu$ branching fractions.  
  However, the second lesson from our study
 is that $1/m_c$ corrections may play an important role, and a further study 
 of their effect is wanted. 
 
 The most serious caveat to our present derivation of a narrow resonance 
 dominance comes from the fact that we have neglected radiative
 corrections. A priori we expect radiative corrections to provide only corrections
 and our present estimate to yield the general trend. This is unhappily not always
 true. As a counterexample see the discussion which follows eq.~(7.8) in
 \cite{bgm}. It is argued that some radiative corrections to the parameter $K$ are
 parametrically larger than the $\alpha_s=0$ estimate. A careful study of 
 radiative corrections to our present sum rule and its consequences would be
 welcome.
 
 It is not excluded that an important 
 fraction of the $B\to D^{(\ast)} \pi
 l \nu$ decays observed at LEP are non-resonant. Unluckily   
 theoretical works addressing non-resonant decays are rare, \cite{gr}
 find in the soft-pion domain a resonance dominance while Isgur \cite{isgur2}
 predicts that no more than 5 \% of the semileptonic decay is  non-resonant. 
  Furthermore, if such a
 continuum contributes significantly, it should also be included in 
 the sum rules \cite{isgur2} and we might fear that at the end 
 of the day the paradox would still be there.  
 
 Finally another experimental result \cite{CLEO} seems to contradict
 our theoretical expectation: the branching ratio for 
 $B\to D_1(j=1/2)\pi^-$ is found to be $\simeq 1.5$ times larger than that of
 $B\to D_1(2420)\pi^-$. Of course the experimental error is still large, and the
 relation between nonleptonic decays and the semileptonic ones assumes 
 factorisation. 
 
 But still {\it there is a puzzle}: on one side an increasing amount of
 theoretical evidence in favor of the narrow resonances dominance, and 
 on the other side an increasing amount of experimental evidence 
 in the opposite direction!

\section{Conclusion and outlook}

We have explicitly checked  quark-hadron duality in the SV limit 
to order $\delta m \Lambda/m_b^2$ including ground state final hadrons and $L=1$
orbitally excited states. We have shown that this duality implied a new sum rule 
eq.~(\ref{new}) which we have also demonstrated from OPE applied to T-product of
axial currents. 

We have shown that this sum rule combined with some theoretical estimates of
$\xi_3$ lead to the conclusion that very probably the $B$ decay into narrow 
$L=1$ resonances was dominant over the one into broad
resonances. This remark seems to contradict recent experimental claims that 
the broad resonances dominate. We have discussed this situation which 
needs urgently further theoretical and experimental work.  

Beyond understanding this experimental puzzle, further theoretical work is needed. 
 For example we might wonder if some proof of the new sum rule
 along the line of \cite{bjorken} is possible. Some progress has been done in
 this direction. The effect of radiative corrections should also be studied.
 
Last but not least, other new sum rules derived
along the same line with other  currents or other components of the currents
 should be considered. 

\acknowledgements 
We thank Patrick Roudeau for many stimulating discussions which have triggered
this study.


\begin{thebibliography}{30}
\bibitem{ac2m2}G. Altarelli, N. Cabibbo, G. Corbo, L. Maiani and G. Martinelli,
Nucl. Phys. {\bf B208} (1982) 365. 
\bibitem{cgg} J. Chay, H. Georgi, B. Grinstein,  
Phys. Lett.  {\bf B247} (1990) 399. 
\bibitem{bsuv93} I. Bigi, M. Shifman, N. Uraltsev, A. Vainshtein, 
Phys. Rev. Lett. {\bf 71} (1993) 496. 
\bibitem{mw} A. Manohar and M. Wise, Phys. Rev.  {\bf D49} (1994) 1310. 
\bibitem{fls} A. Falk, M. Luke, and M. Savage, Phys. Rev. 
 {\bf D49} (1994) 3367. 
\bibitem{shifman} A comprehensive discussion of duality in heavy meson
decays can be found in 
B. Chibisov, R. Dikeman, M. Shifman, N. Uraltsev, Int. J. Mod. Phys. 
A {\bf 12} (1997) 2075; 
B. Blok, M. Shifman, D.-X. Zhang, Phys. Rev. D {\bf 57} (1998) 2691.
\bibitem{wilson}  K.G. Wilson, Phys. Rev. {\bf 179} (1969) 1499; 
Phys. Rept. {\bf 12} (1974) 75. 
\bibitem{luke} M. Luke, Phys. Lett.  {\bf B252} (1990) 447.
\bibitem{isgur} N. Isgur, Phys. Lett. {\bf B448} (1999) 111. 
\bibitem{nous} A. Le Yaouanc, D. Melikhov, 
V. Morenas, L. Oliver, O. P\`ene and J.-C. Raynal, in preparation.
\bibitem{thooft} G. 't Hooft, Nucl. Phys. {\bf B75} (1974) 461.
\bibitem{gl} B. Grinstein and R. Lebed, 
Phys. Rev.  {\bf D57} (1998) 1366.
\bibitem{b-thooft} I. Bigi, M. Shifman, N. Uraltsev and A. Vainshtein, 
Phys. Rev.  {\bf D59} (1999) 054011.
\bibitem{bjorken} J. D. Bjorken, Talk at Les Rencontres de Physique de la 
Valee d'Aoste, La Thuile, Italy, Report No. SLAC-PUB-5278 (1990), 
unpublished; 
J.D. Bjorken, I. Dunietz and J. Taron, Nucl. Phys. {\bf B371} (1992) 111. 
\bibitem{iw}N. Isgur and M. Wise, Phys. Rev.  {\bf D43}, 819 (1991). 
\bibitem{voloshin} M. Voloshin, Phys. Rev.  {\bf D46} (1992) 3062.
\bibitem{greg} A.G. Grozin and G.P. Korchemsky, 
Phys. Rev. {\bf D53} (1996) 1378.
\bibitem{bgm} G. Boyd, B. Grinstein and A. Manohar, 
Phys. Rev. {\bf D54} (1996) 2081.
\bibitem{sv} M. B. Voloshin and M. A. Shifman, 
Sov. J. Nucl. Phys. {\bf 47} (1988) 511. 
\bibitem{llsw}A.K. Leibovich, Z. Ligeti, I.W. Stewart and M.B. Wise,
Phys. Rev. {\bf D57} (1998) 308.
\bibitem{blrw} C.G. Boyd, Z. Ligeti, I.Z. Rothstein and M.B. Wise,
Phys. Rev. {\bf D55} (1997) 3027.
\bibitem{blok}B. Blok, l. Koyrakh, M. Shifman and A.I. Vainshtein
Phys. Rev {\bf D49} (1994) 3356.
\bibitem{fn} A.F. Falk and M. Neubert, Phys. Rev. {\bf D47} (1993) 2965.
\bibitem{bigi} I. Bigi, M. Shifman, N. G. Uraltsev and A. Vainshtein
 Phys.Rev. {\bf D52} (1995) 196.
\bibitem{xi3}M. Neubert, Phys. Rev. {\bf D46} (1992) 3914; 
Z. Ligeti, Y Nir and M. Neubert, Phys. Rev. {\bf D49} (1994) 1302.
\bibitem{dmitri}D. Melikhov, Phys. Rev. {\bf D56} (1997) 7089.
\bibitem{vincent}V. Mor\'enas, A. Le Yaouanc, L. Oliver, O. P\`ene, J.-C. Raynal
Phys. Rev. {\bf D56} (1997) 5668; V. Morenas, th\`ese \`a 
l'Universit\'e Blaise Pascal, Clermont-Ferrand, (1997).
\bibitem{deandrea} A. Deandrea
Presented at 29th International Conference on High-Energy Physics 
(ICHEP 98), Vancouver, Canada,
 23-29 Jul 1998. In Vancouver 1998, High energy physics, vol. 2, 1179.
\bibitem{ebert}D. Ebert, R.N. Faustov and V.O. Galkin,
Talk given at 10th International Seminar on High-Energy Physics
(Quarks 98), Suzdal, Russia, 18-24, May 1998. 
\bibitem{col}P. Colangelo, F. De Fazio and N. Paver,
Talk given at International Euroconference on Quantum Chromodynamics (QCD98),
Montpellier, France, 2 - 8 Jul 1998, Nucl. Phys. Proc. Suppl. {\bf 74} (1999) 222.
\bibitem{yuan} Yuan-ben Dai and Ming-qiu Huang,  Phys. Rev. {\bf D59}
 (1999) 034018.
\bibitem{aleph} D. Buskulic et al., ALEPH, Z. Phys. {\bf C73} (1997) 601.
\bibitem{delphi}P. Abreu et al., DELPHI, CERN-EP/99-174.
\bibitem{gr} J. L. Goity and W. Roberts, Phys. Rev. {\bf D51} (1995) 3459.
\bibitem{isgur2}N. Isgur, Phys. Rev. {\bf D60} (1999) 074030.
\bibitem{CLEO} S. Anderson et al, CLEO Collaboration, hep-ex/9908009.
\end{thebibliography}
\end{document}